# Effective sampling of random surfaces by baby universe surgery

J. Ambjørn

The Niels Bohr Institute
Blegdamsvej 17, DK-2100 Copenhagen Ø, Denmark

P. Białas

Inst. of Comp. Science, Jagellonian University,
ul. Nawojki 11, PL 30-072, Kraków Poland

J. Jurkiewicz

Inst. of Phys., Jagellonian University.,
ul. Reymonta 4, PL-30 059, Kraków 16, Poland

Z. Burda[1][2] and B. Petersson

Fakultät für Physik, Universität Bielefeld,
Postfach 10 01 31, Bielefeld 33501, Germany

## Abstract

We propose a new, very efficient algorithm for sampling of random surfaces in the Monte Carlo simulations, based on so-called baby universe surgery, i.e. cutting and pasting of baby universes. It drastically reduces slowing down as compared to the standard local flip algorithm, thereby allowing simulations of large random surfaces coupled to matter fields. As an example we investigate the efficiency of the algorithm for 2d simplicial gravity interacting with a one-component free scalar field. The radius of gyration is the slowest mode in the standard local flip/shift algorithm. The use of baby universe surgery decreases the autocorrelation time by three order of magnitude for a random surface of $0.5 \cdot 10^5$ triangles, where it is found to be $\tau_{int} = 150 \pm 31$ sweeps.

---

[1]A fellow of the Alexander von Humboldt Foundation.

[2]Permanent address: Inst. of Phys., Jagellonian University., ul. Reymonta 4, PL-30 059, Kraków 16, Poland

# 1 Introduction

Random surfaces play an important role in many branches of physics. They naturally appear in context of membranes, string theory, 2d gravity, QCD strings, dynamics of Nielsen–Olesen vortex, 3d Ising model and many other fields. The basic concept in the theory of random surfaces is the measure of integration over geometries. Two successful methods of functional integration over geometries of random surfaces exist : the continuum approach proposed by Polyakov [1] which leads to the Liouville theory, and the discrete approach, based on dynamical triangulations [5]-[7]. The former is solved using conformal field theory [2]-[4]. It gives predictions for the critical exponent $\gamma$ of the entropy of surfaces, embedded in $d \leq 1$ dimensions, or equivalently for conformal fields with a central charge $c = d$ minimally coupled to gravity. For dimensions $d = c > 1$ the exponent $\gamma$, as well as the critical exponent for the matter field sector, gets an imaginary part and the results have no direct physical meaning. It has not yet been understood if the breakdown at $c = d = 1$ has it's origin in the method itself or whether it reflects some drastic change in the surface entropy, which cannot be described by the exponent $\gamma$ [8]. The discrete approach also allows for analytical solutions for $c \leq 1$, using the equivalence with matrix models. In all solved cases the critical behaviour agrees with that predicted by the continuum conformal field theory. For $c = d > 1$ the discretized models can not yet be solved analytically, but they are perfectly well defined and one can study the models by numerical methods.

In fact, in the past years many attempts have been undertaken to simulate randomly triangulated surfaces by computer. In the standard approach the change in the surface geometry is obtained with the help of a local move (called *flip*), described for instance in [7] and [9]. The biggest problem connected with the numerical study of the critical properties of triangulated systems, particularly for $d > 1$, is the critical slowing down, which restricts the range of simulations to rather small lattices. Already for lattices of a moderate size it is difficult to thermalize the system and then generate independent samples, by the use of this standard algorithm. The reason of the slowing down can be traced to the fact that the local algorithm, although ergodic, is not well suited to update the typical geometry known to be dominated by the nonlocal structures of *baby universes* or branched polymers, for which the entropy coming from rearranging whole sub–universes seems to play an important role for the effective picture of geometrical fluctuations [10]-[19].

Since the standard algorithm was proposed [7],[9], not many improvements have been made. Quantitatively new algorithms were proposed in [10, 11]. The algorithm in [10] generates independent triangulations for $d = 0$ (pure gravity) making use of the Dyson–Schwinger equations which determine relations between graphs. These relations are rewritten in Monte–Carlo language and used to sample sub–universes in a recursive way. Roughly speaking, the method is based on the theoretical input coming from exact formulas for the distribution $n(A, l)$ of sub–universes with area $A$ and perimeter $l$. The relations between $n(A, l)$ are known only for pure gravity and the method is limited to this case. Similar remarks are valid for the algorithm suggested in [11]. It is specific to $d = -2$.

Using "cluster" algorithms [12] can reduce very much the critical slowing down in the updating of the matter field sector. This algorithms were succesfully used to study spin systems interacting with two– and four–dimensional simplicial gravity [13],[14]. In [15] an algorithm called "valleys-to-mountains reflections" was proposed to reduce the large correlation time in the updating of the continuous matter field variables. This algorithm was used in [16] to study the case $d = 1$. As was claimed there the autocorrelation times were of the order of 300 sweeps for a system with 30.000 triangles. It is not completely clear to us how this analysis was made. Our experience with "cluster" algorithms shows that even a very fast algorithm in the matter sector does not help to improve the updating of geometry and in effect we would expect the autocorrelation time in this case to be much longer.

In this letter we propose a new, very general and efficient updating scheme. In this algorithm, apart from the standard sweeps of the lattice, using a local *flip* move, we introduce a new type of move, which we call a *big move* or *baby universe surgery*. The *big move* is a generalization of the Alexander moves [17], discussed recently in the context of higher dimensional simplicial gravity [23]. The idea is to introduce large changes in the geometry, typical for structures resembling branched polymers, which at the same time have large acceptance rate.

As we will show, the autocorrelation time for the new update is dramatically decreased compared to the standard, local algorithm. For a lattice with $0.5 \cdot 10^5$ triangles we found an autocorrelation time of $150 \pm 31$ sweeps for the slowest mode. This should be compared with the correlation time obtained using the standard algorithm, where we already for a lattice with 2396 triangles observed an autocorrelation time of $1900 \pm 512$.

## 2 Algorithm

Recently, the fractal structure of 2d gravity was described by the distribution of so-called baby universes [18], which are sub–universes with relatively large area $A$, connected with the *mother universe* through loops with small perimeter $l$. Their distribution determines the leading term in the entropy of surfaces, a fact which has already been used successfully in numerical simulations to determine $\gamma$ for some quantum gravity models [24, 25, 26]. These sub–universes will be the main object in our updating scheme. More precisely, we shall concentrate on the minimal neck baby universes, called *minbus*.

Let us first describe the *big move* algorithm for the case of pure gravity and for surfaces with a spherical topology. In the first step of the algorithm, we find a minimal neck on a surface, *i.e.* a loop which has three links and divides the surface into two parts. The smaller part will be called *minbu* and is the main object in our updating scheme. The larger part will be called the *mother universe*. The minimal neck is a triangle, which does not belong to the surface (see fig. 1). If we cut the surface across this neck and fill the triangular holes on both sides of the cut, the sphere splits into two surfaces, both with a topology of a sphere. These two surfaces can now be glued back in a different way. To do this we chose randomly one triangle on each surface and remove these triangles, changing the two spherical surfaces into discs with triangular edges. These edges are glued together, forming a minimal neck of the new surface. This operation, which we denote *baby universe surgery*, can be performed in six different ways, depending on the way the vertices of the triangles are to be identified. This identification is chosen at random. It can be easily seen that such a *big move* preserves the total area of the surface. For the pure gravity case the detailed balance condition is automatically satisfied, because the move is reversible and all the triangulations have the same weight. The area of the minbu involved in the move can be quite large. For surfaces with the polymer structure the move can be visualized as cutting of a branch and gluing it back in a random way. Performing such a move with the help of the local *flips* would require very long computation time.

The move we use in this paper is a slightly simplified version of the one described above. The simplification lies in the fact that we choose a new triangle only on the bigger surface and keep the position of the minimal neck on the *minbu* in the terminology introduced above unchanged. The *big moves* are supplemented with the standard sweeps of the lattice, making use of the local *flips*.

The concept of a *big move* can be generalized to the case of gravity coupled to matter fields. As an example consider a $d$–component free bosonic field on a randomly triangulated surface. In the discretized approach we choose a version, where the field is located in the middle of the triangles. The field configuration is represented by a set of real numbers $x_i^\mu$, $\mu = 1, \ldots d$, where $i$ labels the triangles of the surface. The action of the field is

$$S = \sum_{ij,\mu} (x_i^\mu - x_j^\mu)^2, \tag{1}$$

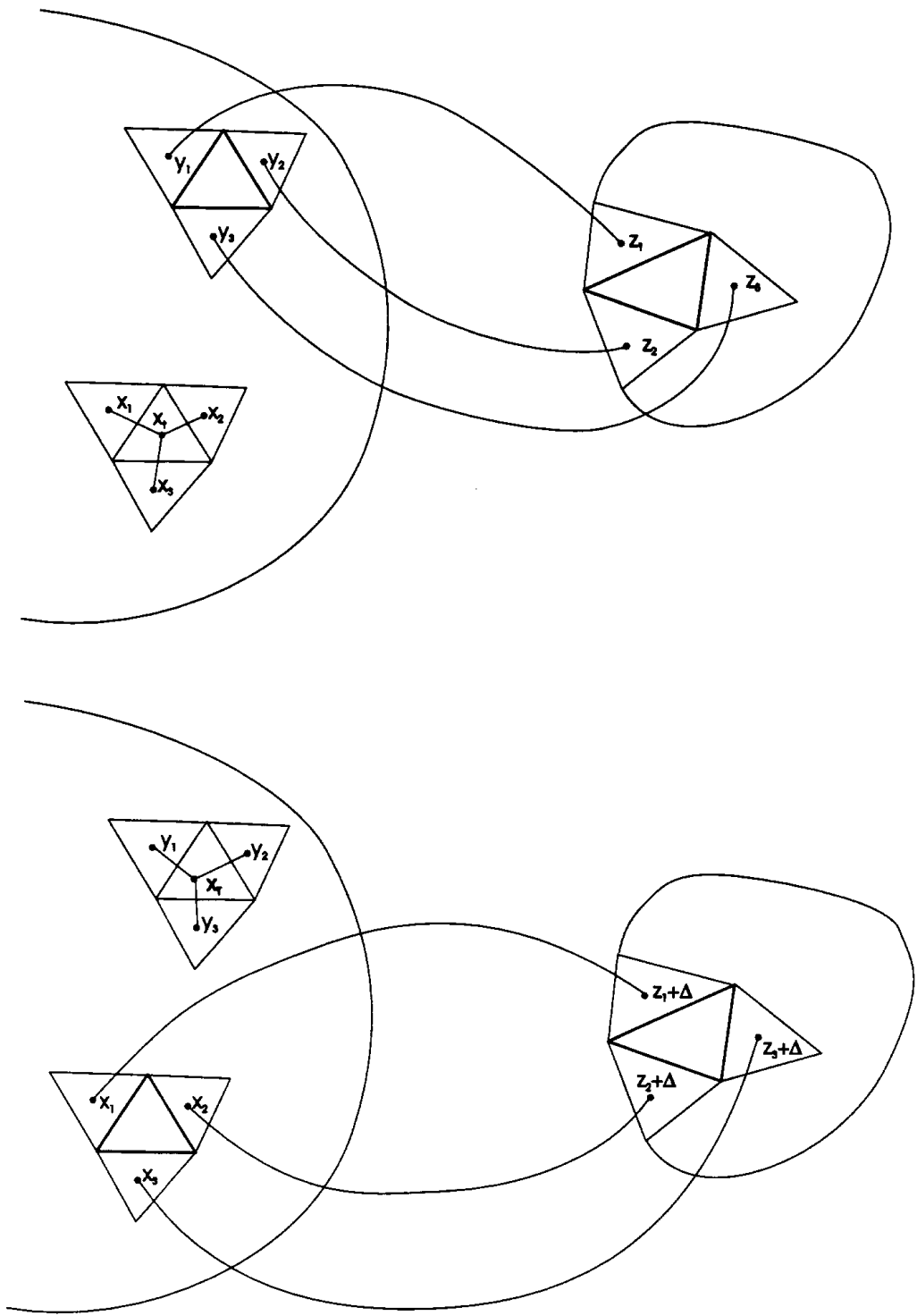

Figure 1: The *big move* of the surface. The situation before and after the move is displayed.

where the sum runs over all pairs of the neighboring triangles. In this case, the naive replacement of the position of the minbu can lead to a large change of the action and very small acceptance rate. To overcome this problem, while performing the replacement one has to propose a new $x$–field configuration. In order to maximize the acceptance rate for the transition one should minimize possible changes of the matter field. It can be realized by keeping fixed most of the squares of the relative differences between fields $x_i^\mu$ on the *minbu* and on the *mother universe*, so that the only change of the action comes from the change of interactions between the $x$ fields nearest to the loops on which the algorithm cuts and pastes a *minbu*.

We propose to change all fields on the minbu by adding a constant shift $\Delta^\mu$ to them :
$$x^\mu \to x^\mu + \Delta^\mu. \qquad (2)$$

When performing a *big move*, a new field $x_T^\mu$ has to be created in the center of the new triangle $T$ on the bigger surface, obtained in place of the minimal neck. At the same time the field $x_t^\mu$ in the center of the triangle $t$, which becomes a new minimal neck has to disappear. Altogether there are $2d$ numbers ($d$ for $\Delta^\mu$ and $d$ for $x_T^\mu$) to specify completely a transition between the configurations, and therefore it is clear that the simple Metropolis question imposed on $2d$ randomly chosen fields, is ruled out, since one would get very low acceptance. We found several possible solutions of how to update the field sector effectively. The most efficient one, which we describe below, is a version of the heat–bath algorithm. We make the update in two steps. First we ask, if the replacement of the minbu and triangle is accepted. We do not specify $x_T^\mu$ and $\Delta^\mu$ but instead integrate over them. This gives us the volume of the new available state space. To satisfy a detailed balance condition the transition must be reversible, and therefore this volume has to be compared with the analogous volume integrated over $x_t^\mu, \Delta^\mu$ for the inverse transition. Denote the volumes for the configurations $A$, $B$, before and after transition by $V_A$ and $V_B$. The detailed balance for this transition reads :
$$V_A \; p(A \to B) = V_B \; p(B \to A). \qquad (3)$$

To compute the volumes, denote the fields around the minimal neck on the *minbu* by $z_1, z_2, z_3$ and on the bigger surface by $y_1, y_2, y_3$ (see fig. 1). In the following the index $\mu$ will be omitted. The fields $y_1, y_2, y_3$ interact with the new field $x_T$ in the center of the triangle, yielding the volume in the state space :
$$\begin{aligned} I(y_1,y_2,y_3) &= \int d^d x_T \exp\{-\sum_{i=1}^{3}(y_i - x_T)^2\} \qquad (4) \\ &= \mathcal{N} \exp\{3(\langle y^2 \rangle - \langle y \rangle^2)\} \end{aligned}$$

Let us denote the fields around the triangle $t$ as $x_i$. After gluing the fields $x_i$ interact with their counterparts $z_i$ on the *minbu*. According to our prescription the whole *minbu* can be shifted by $\Delta$ : $z_i \to z_i + \Delta$. The state volume is obtained by integrating over $\Delta$:

$$\begin{align}
J(x_1, x_2, x_3; z_1, z_2, z_3) &= \int d^d\Delta \exp\{-\sum_{i=1}^{3}(x_i - z_i + \Delta)^2\} \tag{5}\\
&= \mathcal{N} \exp\{3(\langle(x-z)^2\rangle - \langle(x-z)\rangle^2)\}\\
&= I(x_1 - z_1, x_2 - z_2, x_3 - z_3).
\end{align}$$

The detailed balance condition can be obtained by the comparison of this move with the inverse move. It has the form :

$$I(y)I(x-z)p(A \to B) = I(x)I(y-z)p(B \to A) \tag{6}$$

and is satisfied by the probability $p(A \to B)$ in the form :

$$\begin{align}
p(A \to B) &= \max\left\{1, \tfrac{I(y-z)}{I(y)I(z)} \Big/ \tfrac{I(x-z)}{I(x)I(z)}\right\}\\
&= \max\left\{1, \tfrac{\exp 6(\langle yz\rangle - \langle y\rangle\langle z\rangle)}{\exp 6(\langle xz\rangle - \langle x\rangle\langle z\rangle)}\right\},
\end{align} \tag{7}$$

where

$$\langle xz\rangle - \langle x\rangle\langle z\rangle) = \frac{1}{3}\sum_{i,\mu} x_i^\mu z_i^\mu - \frac{1}{9}\sum_{i,j,\mu} x_i^\mu z_j^\mu. \tag{8}$$

If the cut/paste move is accepted, the next step is to assign $x_T$ and $\Delta$ with the appropriate gaussian distributions

$$\propto \frac{1}{\mathcal{N}} d^d x_T \exp(-3(x_T - \langle x\rangle)^2) \tag{9}$$

for $x_T$ and

$$\propto \frac{1}{\mathcal{N}} d^d\Delta \exp(-3(\Delta + \langle y - z\rangle)^2) \tag{10}$$

for $\Delta$, with $\mathcal{N} = (\pi/3)^{3/2}$, as in (5) and (6).

Notice, that the transition probability (7) has to be modified if triangles $T$ and $t$ have a common link.

The algorithm can easily be generalized to surfaces with higher genus. Other matter fields can also be introduced.

## 3   Numerical simulations

To check the performance of our algorithm we performed numerical simulations of a two dimensional random surface interacting with a one–component scalar field. This system corresponds to the boundary case $c = d = 1$ and we expect the algorithm to perform even better for $c = d > 1$.

In our simulations we concentrated on three types of observables important for the effective picture of a fluctuating surface : observables reflecting

(a) short range structure of the surface,

(b) a long range, global geometric structure and

(c) the matter sector, which through it's coupling influences the internal geometry of the surface.

These three types of observables are in the standard local updating scheme characterized by three different time scales of evolution, described by the corresponding autocorrelation times. We present here results for only few observables which we found representative for each type of observables. As a example of a type (a) observable we consider the average square of curvature which is the discretized version of :

$$\langle R^2 \rangle = \int \sqrt{g} R^2 / \int \sqrt{g}. \tag{11}$$

A type (b) observable is the geodetic distance $d_{xy}$ between two points on the surface with fixed labels $x$ and $y$. Another quantity of this type is the average internal surface extension *i.e.* the distance averaged over all pairs of points

$$\bar{d} = \langle d_{xy} \rangle. \tag{12}$$

The lattice is invariant with respect to the permutations of the point indices, which are in fact only dummy arguments. It means that $d_{xy}$ and $\bar{d}$ estimate the same quantity. In the updating procedure the endpoints $x, y$ perform random movements over the lattice and after a long measurement time $d_{xy}$ should equal $\bar{d}$, but with much bigger error. The reason we study them independently is that we are mainly interested in the algorithm dynamics, and the two show different behaviour during the updating procedure. As a type (c) observable we choose the gyration radius $r$

$$r^2 = \langle (x - \bar{x})^2 \rangle. \tag{13}$$

To compare the autocorrelation times we performed runs using two types of algorithm. For both types the time was measured in *sweeps*. The first algorithm, which we call the *local* algorithm, made use only of the local moves. The geometry was updated using the *flip* moves. A sweep consists of the number of attempted flips equal to the number of lattice links, followed by a shift, which is the heat bath and overrelaxation update of all the $x$'s. In practical calculations the overrelaxation was used with a probability 50%, which we found to minimize the autocorrelations for the local algorithm. The second algorithm, which we call the *global* algorithm, used the local sweeps described above together with the global sweeps consisting of the *big moves*, where the *big move* was attempted at each minimal neck of the surface. For the *global* algorithm a sweep means either the local or the global sweep, each performed at random with equal probability.

Table 1: The integrated autoccorelation times for the local and global algorithms.

| $N_T$ | $\tau_{int}(R^2)$ | | $\tau_{int}(d_{xy})$ | | $\tau_{int}(d)$ | | $\tau_{int}(r^2)$ | |
|---|---|---|---|---|---|---|---|---|
| | local | global | local | global | local | global | local | global |
| 46 | 1.18(2) | 1.66(3) | 5.3(1) | 2.61(7) | 1.01(2) | 1.20(2) | 9.5(4) | 3.26(9) |
| 76 | 1.60(3) | 1.92(4) | 7.9(3) | 3.04(8) | 1.87(4) | 1.70(4) | 24(2) | 4.2(1) |
| 116 | 1.97(4) | 2.13(5) | 11.5(5) | 3.08(9) | 2.85(8) | 2.23(5) | 38(3) | 5.1(1) |
| 156 | 2.12(5) | 2.42(6) | 14.7(8) | 3.22(9) | 4.2(1) | 2.85(8) | 63(7) | 5.7(2) |
| 236 | 2.36(6) | 2.57(7) | 21(1) | 3.3(1) | 5.7(2) | 4.0(1) | 86(11) | 7.3(2) |
| 316 | 2.57(6) | 3.01(8) | 26(2) | 3.4(1) | 8.1(3) | 5.1(1) | 116(18) | 8.9(3) |
| 396 | 2.63(7) | 2.91(8) | 30(2) | 3.6(1) | 11.4(5) | 5.7(2) | 201(41) | 8.6(3) |
| 796 | 2.81(8) | 3.5(1) | 53(6) | 4.0(1) | 24(2) | 10.2(4) | 430(126) | 13.3(7) |
| 1596 | 3.04(6) | 3.53(7) | 92(9) | 4.3(1) | 40(3) | 19.5(8) | 1118(374) | 20.4(9) |
| 2396 | 2.99(3) | 3.7(1) | 97(6) | 4.3(2) | 39(2) | 23(2) | 1900(512) | 25(2) |
| 49996 | | 4.6(2) | | 9.2(5) | | 241(65) | | 151(31) |

For each observable we measured the integrated autocorrelation time and fitted it to the asymptotic formula

$$\tau_{int} = cA^z, \tag{14}$$

to extract the dynamical exponent $z$. In (14), $A$ is the surface area, which is equal to the number of triangles. The fit was made using the standard MINUIT program library [28]. The exponent $z$ can be different for different quantities and the real autocorrelation time can be identified with that of the slowest mode. In our simulations we covered the range of sizes up to 2400 triangles for the *local* algorithm and up to $0.5 \cdot 10^5$ for the *global* algorithm. Effective simulations with lattices of that size was possible thanks to a very large reduction of the autocorrelation times.

The measured autocorrelation times (in sweeps) are presented in the table 1 both for the local and the global algorithm. In the first column we show the size dependence of the autocorrelation time $\tau_{int}$ for $\langle R^2 \rangle$. For this quantity $\tau_{int}$ scales very slowly with the lattice size. The average curvature square is the fastest mode in the dynamics of both algorithms. From the table it is seen that for local quantities there is no gain in the autocorrelation time from the *big moves*. The *flips* decorrelate the local geometrical quantities faster than *big moves*. In fact including the *big moves* makes the autocorrelation time slightly longer. This effect is probably related to the reduction of the autocorrelation time for the matter sector, described below. It is also clear that in both algorithms the autocorrelations in this sector are short–ranged and have no influence on the real autocorrelation time.

In the second column we present the results for the integrated autocorrelation time for $d_{xy}$. The *global* algorithm, as compared to the standard *local* one, reduces considerably the autocorrelation time of this observable. It is clear that by cutting and pasting a minbu one changes the branch structure of the universe and in effect a distance between points $x$ and $y$, lying on different sub–universes, can change very much. In this way $d_{xy}$ gets easily decorrelated. The *flips* need much more time for this. The reduction of the autocorrelation time is reflected in the exponent $z$ which we find to be $z = 0.81(6)$ for the local algorithm, and $z = 0.14(2)$ for the global one. To get some idea about the performance of both algorithms for larger lattices, we take as a reference point a lattice of the size $0.5 \cdot 10^5$, which we simulated by the new algorithm. If one extrapolates for the *local* algorithm the results from smaller lattice sizes, one gets $\tau_{int}$ of the order $10^3$-$10^4$, which is two, three orders of magnitude larger than $9.2 \pm .5$ which we got from simulations with the new algorithm.

The relatively short autocorrelation time observed for $d_{xy}$ is however not a good estimate of the correlations for the long–range observables. Since fixing point labels is not physical, as was discussed above, it can be viewed rather as a measure of the mobility of the algorithm. The more realistic estimate of these correlations can be obtained by studying the observable $\bar{d}$. In fact numerically the two estimates are equal within errors. The third column of the table 1 shows that in this case the autocorrelation time grows faster with volume than for the observable $d_{xy}$. There is again a large reduction of the autocorrelation time for the *global* algorithm, the dynamical exponents we get are respectively $z = 1.06(3)$ and $z = 0.76(3)$. The reason of the large increase of $z$ for the global algorithm, compared to that for $d_{xy}$ can be attributed to the fact, that cutting and pasting branched parts of the universe does not change its size too much. Some of the changes done by moving sub–universes are probably undone by next moves. We hope however, that this exponent can be reduced by properly adjusting the ratio of local and global updates, similarly as is the case for the local overrelaxation in the standard algorithms, where $z$ is drastically reduced only if overrelaxation is applied with a proper frequency. From simulations on the lattice with a size $0.5 \cdot 10^5$ we got $\tau_{int} = 241. \pm 65.$ and it is two orders of magnitude lower than that obtained from extrapolating the results for the local algorithm.

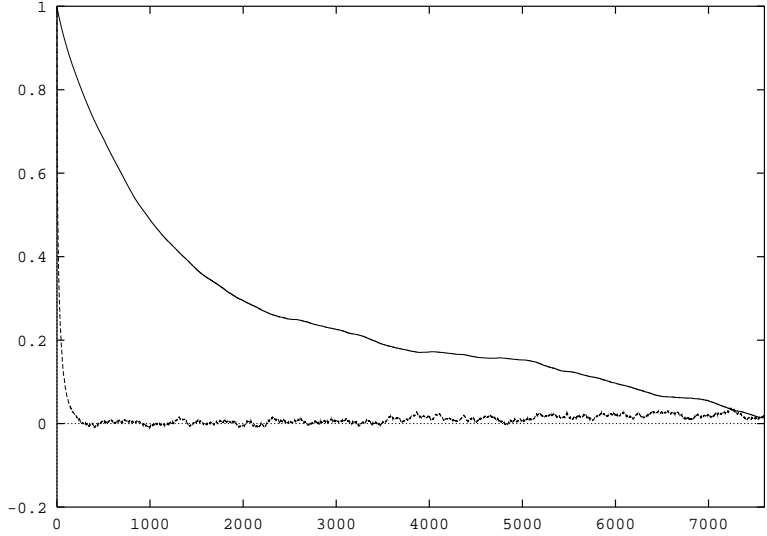

Figure 2: The normalized autocorrelation functions for gyration radius on the lattice with 2396 points.

The real problem for the *local algorithm* is created by the slowest mode which is the matter sector. In this sector for the standard local algorithm we find the dynamical exponent $z = 1.4 \pm 0.1$ for the gyration radius $r^2$. The extrapolation of the results from small lattices up to the surface with $0.5 \cdot 10^5$ triangles gives the number of $10^5$ sweeps needed to decorrelate configurations. In terms of HP 720 CPU computer time it would mean that one needs roughly one week to produce two independent configurations. Using our algorithm we managed to reduce this time to $150. \pm 31.$ sweeps and the exponent $z = 0.50 \pm 0.03$. Already for small volumes the jump in performance makes a real difference between the two algorithms, which is visualized in the figure 2, where the normalized autocorrelation functions for the gyration radius are depicted, showing drastic change of the correlation range.

The results for the dynamical exponents $z$ obtained from a fit $\tau_{int} = cA^z$ for different observables, are summarized in the table 2. The errors quoted are obtained using the MINUIT library program. Comparing the slowest modes for the algorithms one finds that the difference in efficiency is governed by the exponent difference $z_{r^2}^{local} - z_d^{global} = 0.6(2)$. Notice that for large lattices in the *global* algorithm, the longest autocorrelation time comes from the observable $\bar{d}$ ($z = 0.76(3)$) in place of $r^2$ ($z = 0.50(3)$) as is shown in the fig. 3. As mentioned earlier, we hope, however, that $z_d$ can be reduced by adjusting properly the frequency of *big moves*, probably scaling it with the lattice size. We postpone this discussion to a further study.

Table 2: The fits of the volume dependence of the autoccorelation times to the formula $\tau_{int} = cA^z$.

| $\tau_{int} = c \cdot A^z$ | $d_{xy}$ | | $\bar{d}$ | | $r^2$ | |
|---|---|---|---|---|---|---|
| | local | global | local | global | local | global |
| $c$ | 0.23(6) | 1.5(2) | 0.017(3) | 0.06(1) | 0.05(3) | 0.46(9) |
| $z$ | 0.81(6) | 0.14(2) | 1.06(3) | 0.76(3) | 1.4(1) | 0.50(3) |
| $\chi^2$/d.o.f. | 0.11 | 2.18 | 2.66 | 1.40 | 1.17 | 1.12 |

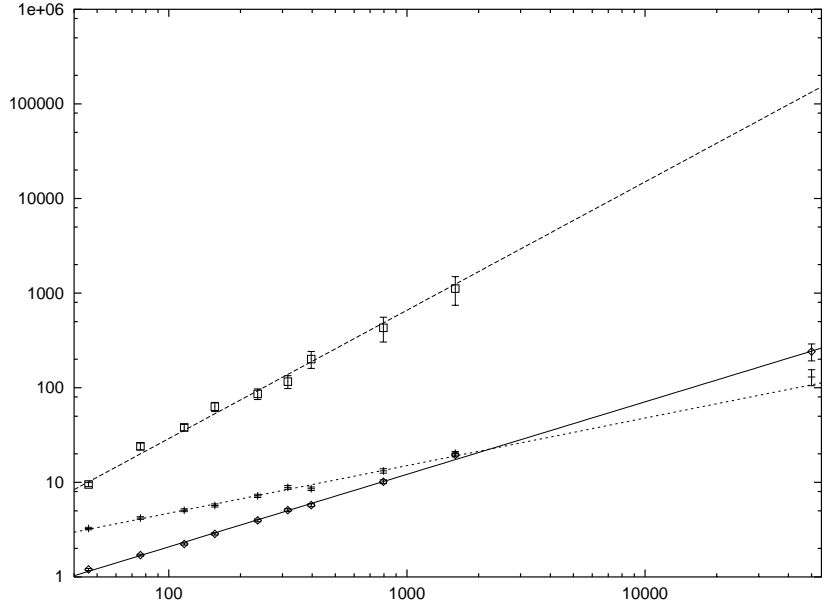

Figure 3: The autocorrelation time *vs.* lattice size. The dashed line corresponds to the autocorrelation time for $r^2$ in the local algorithm which is the slowest mode in this algorithm, the dot–dashed line, the same but in the global algorithm, and the solid line, the autocorrelation time of $\bar{d}$. Notice, that for the lattice size of the order of $10^3$ triangles, it becomes the slowest mode in the new algorithm.

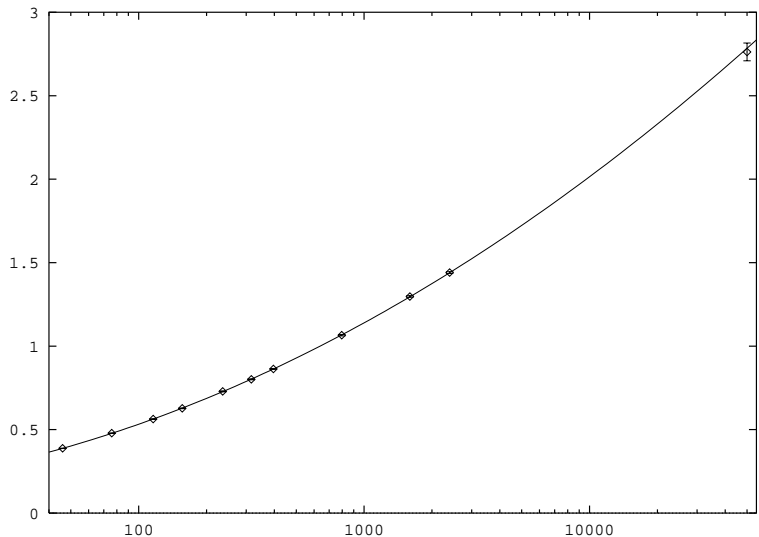

Figure 4: Gyration radius as a function of the lattice size fitted to the formula $r^2 = a + b \log A + c(\log A)^2$. The error bars are smaller than the symbols used for the data points.

From the point of view of the string theory, the gyration radius gives an insight into how the string geometry looks like in the target space in which a string is embedded. For $d = 1$, the theoretical calculations predict for the behaviour of the gyration radius the formula [10],[16] :

$$\langle r^2 \rangle = a + b \log A + c(\log A)^2, \qquad (15)$$

which we verify here numerically using the *global* algorithm. In the fig. 4, $\langle r^2 \rangle$ is plotted *vs.* $\log A$, where $A$ is the number of triangles of the lattice. We find $c = 0.025(2)$, where the error is estimated by comparing the fit to (15) with the one, where the term $\log A/A$ is included. This fit can be compared with the one obtained in [16]. The normalization of the $x$ field we use here is different than in [16]. Their result seems to correspond to $c = 0.020(1)$ in our normalization, although the relative normalizaton of the $x$ field in the two papers is not completely clear to us. Our value seems to be above the one quoted there which may be due to the larger system we use.

## 4  Discussion

The algorithm presented above can be used to study the model of 2$d$ gravity interacting with the Gaussian or spin–like fields for cases when $c \geq 1$. This problem is presently being investigated. Another possible application is in numerical simulations of string models with extrinsic curvature terms in the action [21]. For such models the autocorrelation time for observables like the gyration radius, which are defined in target space, becomes enormous. Even for lattices of the size $12^2$ it is of the order of $10^4$ [20]. This is of course a serious barrier for going to larger volumes, and even for volumes simulated so far, it seems to be a source of debate concerning the interpretation of the results, as for example the order of the phase transition [22]. We hope that the new algorithm will be efficient also in the study of higher dimensional gravity [23, 27], where a cold phase is known to be dominated by elongated branching structure which slows down the standard algorithm based on local decompositions of the simplicial manifold.

As compared to "cluster" algorithms [13],[15],[16], the *big move* algorithm performs large changes of geometry and not only large changes of the matter fields. The algorithm owns its efficiency to the fact that it updates directly the degrees of freedom which seem to be important for the effective picture of typical world–sheet geometry. Incidentally, when performed on a singular spot of a lattice, a *flip* move can drastically change the minbu structure by splitting or gluing some of them together. In general one can think of the flip as the move which is responsible for updating local fluctuations on the surface, like for example the curvature fluctuations. In turn, cutting and pasting minbus is responsible for updating the global branching structure by controlling the part of the entropy which results from baby universe surgery. As shown recently, this type of degrees of freedom can drastically change the total effective entropy of the model [19], and can be very important in the effective picture of the surface. We hope that the new algorithm will play the same role for the random surfaces as the cluster algorithm for spin systems or recursive sampling in $c = 0$ quantum gravity.

**Acknowledgement**
One of us (Z.B.) would like to thank Alexander von Humboldt Foundation for the fellowship. We thank HLRZ-Jülich for the computer time. The work was partially supported by KBN grants : 2P30216904 and 2P30204705.